\documentclass[]{spie}  

\usepackage{amsmath,amsfonts,amssymb}
\usepackage{graphicx}
\usepackage[colorlinks=true, allcolors=blue]{hyperref}

\title{Investigating the pion source function in heavy-ion collisions with the EPOS model}

\author[a]{Maria Stefaniak}
\author[b]{D\'aniel Kincses}
\affil[a]{Warsaw University of Technology, pl. Politechniki 1, 00-661 Warsaw, Poland}
\affil[a]{IMT Atlantique - Subatech, 4 Rue Alfred Kastler, 44300 Nantes, France}
\affil[b]{E\"otv\"os Lor\'and University, P\'azm\'any P\'eter s\'et\'any 1/A, 1117 Budapest, Hungary}
\authorinfo{Further author information: \\D.K.: E-mail: kincses@ttk.elte.hu, \\  M.S.: E-mail: maria.stefaniak.dokt@pw.edu.pl}

\begin{document} 
\maketitle

\begin{abstract}

By measuring the momentum correlations of pions created in heavy-ion collisions we can gain information about the space-time geometry of the particle emitting source. Recent experimental results from multiple different collaborations demonstrated that to properly describe the shape of the measured correlation functions, one needs to go beyond the Gaussian approximation. Some studies suggest that the L\'evy distribution could provide a good description of the source. 
While there are already many experimental results, there is very little input from the phenomenology side in explanation of the observed non-Gaussian source shapes. The EPOS model is a sophisticated hybrid model where the evolution of the newly-created system is governed by Parton-Based Gribov-Regge theory. It has already proved to be successful in describing many different experimental observations for the systems characterized by baryon chemical potential close to zero, but so far the source shape has not been explored in detail. In this paper we discuss studies of the pion emitting source based on the theoretical approach of the EPOS model.
  
\end{abstract}


\section{INTRODUCTION}
\label{s:intro}  

In order to explore the properties of the strongly interacting matter created in high energy heavy-ion collisions one needs to study the space-time geometry of the particle emitting source. The discovery of quantum-statistical (femtoscopic) correlations of particles produced in such collisions facilitated the experimental investigation of the source dynamics~\cite{Goldhaber:1959mj,Goldhaber:1960sf}. Phenomenological studies, e.g. hydrodynamical model calculations substantiated the assumption that the shape of the source is Gaussian~\cite{Csorgo:1994fg,Akkelin:1995gh}, however, experimental results from the field of femtoscopy unveiled details that challenged said assumption. The first signs of non-gaussian behavior came from source imaging studies~\cite{Brown:1997ku,Adler:2006as} -- these results implicated that the two-pion source function has a long-range component, it exhibits a power-law behavior. Recent results from different experiments corroborated these studies, and showed that L\'evy type sources can provide a statistically acceptable description of the measured femtoscopic correlations~\cite{Adare:2017vig}. The emergence of these type of source functions in heavy-ion reactions can have multiple different reasons such as anomalous diffusion, jet fragmentation, critical behavior, or resonances~\cite{Csanad:2007fr,Csorgo:2004sr,Csorgo:2005it}. In order to better understand the underlying physical processes behind the experimental findings, it is important to conduct new phenomenological investigations. Event generators modeling nuclear reactions have the advantage of providing direct access to the freeze-out dynamics, they allow to circumvent the process of working backwards from femtoscopical correlations. One of such event generators is the EPOS model -- the \textbf{E}nergy conserving quantum mechanical multiple scattering approach, based on \textbf{P}artons (parton ladders), \textbf{O}ff-shell remnants, and \textbf{S}plitting of parton ladders~\cite{Werner:2010aa}. 

The structure of the paper is as follows: in Section \ref{s:epos}. we discuss the most important details of the EPOS model. In Section \ref{s:source}. we discuss the basic definitions and details of the two-particle source function. In Section \ref{s:results}. we present the first results of the shape analysis of the two-pion source in the EPOS model. Finally, in Section \ref{s:sum}. we summarize our findings.

\section{The EPOS model}
\label{s:epos}
EPOS is based on Monte Carlo techniques. In the simulations the parton-based Gribov-Regge theory (PBGRT) is considered which is a mixture of two approaches \cite{Drescher:2000ha}. The theoretical framework consists of the coherent description of the space-time expansion of matter based on research of both elementary processes such as lepton-nucleon scattering or electron-positron annihilation and more complex collisions of protons or nuclei. 

\begin{figure}
\centering
\includegraphics[scale=0.4]{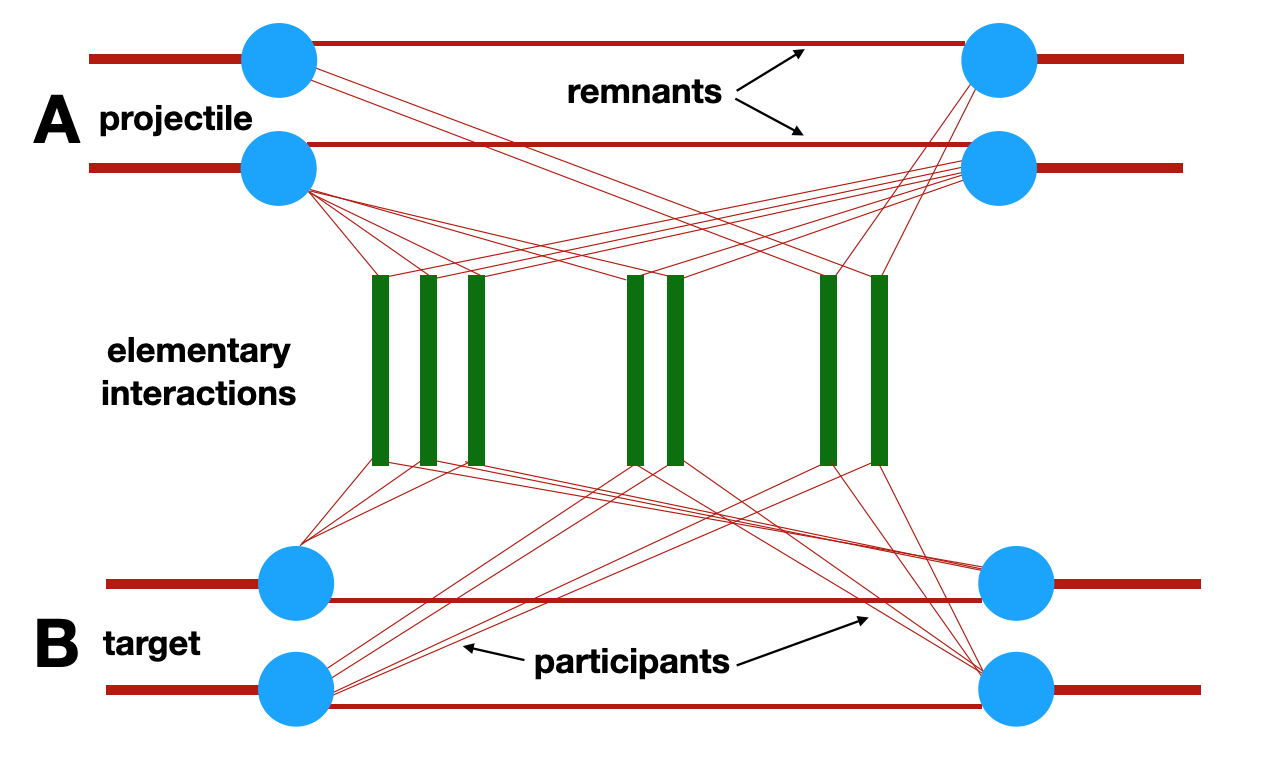}
\caption{Illustration of the nucleon-nucleon rescattering. In the example an interaction of two projectiles (A) and two targets (B) is shown. The split between participants and remnants shows the momentum sharing between constituents ensuring conservation of given valuable.}
\label{fig:remnant}
\end{figure}

The Gribov-Regge, an effective-field theory, provides the description of the soft interactions \cite{Gribov:1968fc}, while the eikonalized parton model introduces partons and gluons. It also gives a recipe how to calculate the parton cross-sections. 
The combination of these two approaches in the initial stage of the evolution ensures the preservation of the conservation laws and equal treatment of subsequent pomerons (corresponding to elementary interactions). In EPOS both particle production and cross-sections are treated by the same formalism. Figure~\ref{fig:remnant}. shows an illustration of the proton-proton and proton-nucleus collisions. The elementary interactions are represented by the green lines. They consists of the soft, semi-hard and hard contributions. The energy of the incoming projectile and the target is shared between participants and remnants. 

\begin{figure}
\centering
\includegraphics[scale=0.5]{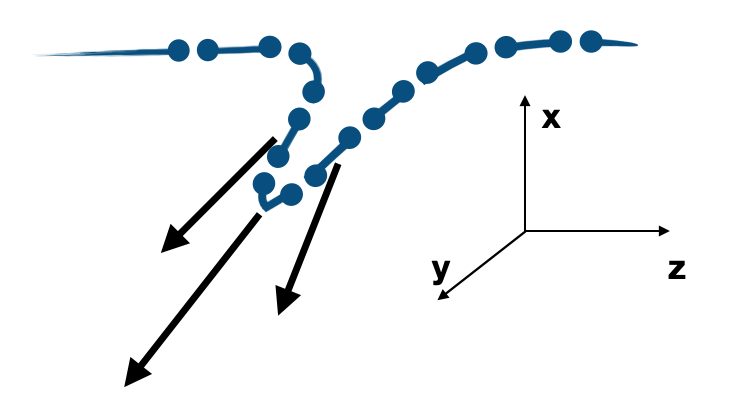}
\caption{The flux tube divided into segments with the kink, moving transversely.}
\label{fig:kink_ft}
\end{figure}

PBGRT introduces open and closed \textit{parton ladders}, which correspond to inelastic scattering (new particles are produced) and elastic scattering, respectively. In EPOS, ladders are expressed by the quasi-longitudinal color field, the elementary flux tubes and recognized as classical strings \cite{Werner:2010aa}. The fragmentation of the tubes occurs according to the String Model approach \cite{Andersson:1986au, Ferreres-Sole:2018vgo}. Strings are divided into multiple segments which are characterized by the parton velocity and length by parton energy. The transverse motion is introduced into the \textit{kinky string} evolution by the intermediate gluons (illustrated in Fig.~\ref{fig:kink_ft}).

In the following subsections we discuss the three main steps of the model, the core-corona division, the hydrodinamical evolution, and the hadronic cascades.

\subsection{Core-corona division}

The crucial part of the evolution of the system in EPOS is the moment of medium division into \textit{core} and \textit{corona}. The separation of the segments originating from the fragmentation of the flux tubes is based on the possibility to escape from the "bulk matter". The main criteria is transverse momentum and the local string density. If the segment is located close to the kink, it is characterized by high transverse momentum and it easily leaves the matter - these become the corona particles and hadronize as a hadron jet. On the other hand, if the segment belongs to the dense area and has low transverse momentum it becomes the core matter, which expands according to the rules of hydrodynamics. There is also a third possibility that the string segment is located close to the surface of the dense fraction of medium and its transverse momentum is high enough to escape - consequently it becomes a corona particle. The calculations needed for the division are performed according to the following equation:
\begin{equation}
    p^{new}_{t} = p_{t} - f_{Eloss}\int_{\gamma} \rho dL,
\end{equation}
where $\gamma$ is the trajectory of the segment, $\rho$ is the density, $f_{Eloss}\neq 0$ constant for $p_{t}<p_{t,1}$,  $f_{Eloss}= 0$ for $p_{t}>p_{t,2}$. If $p^{new}_{t}$ is negative the segment becomes a part of the core medium, otherwise it contributes to corona \cite{Werner:2013tya, Werner:2010aa, Werner:2007bf}.

\subsection{Hydrodynamical evolution}

The model consist of the 3D+1 viscous hydrodynamics called vHLLE (viscous HLLE-based algorithm). It is based on a realistic Equation of State, which is compatible with Lattice QCD data \cite{Allton:2002zi}.
In EPOS each event is treated as an individual case, the generalizations of smoothing initial conditions for all simulated events is not considered. In the hydrodynamical evolution, an event-by-event (\textit{ebe}) approach is introduced based on the randomization of the flux tube initial conditions \cite{Werner:2010aa}. It is significant in the studies of spectra and flow harmonics. The hadronization process is based on the Cooper-Frye procedure, where the hyper surface is characterized by constant energy density $\varepsilon = 0.57$ GeV/fm$^{3}$.

\subsection{Hadronic Cascades}

The final part of the evolution of the system in EPOS consists of the \textit{hadronic afterburner} - UrQMD \cite{Bleicher:1999xi,Bass:1998ca}. With the cooling of the system, it dilutes and when it reaches low enough density the kinetic theory can be applied. The interactions of particles are only possible from the moment of escape from the hadronization hyper surface. There are 60 various baryonic species and corresponding anti-particles, and about 40 mesonic states treated by the final rescattering. The interactions between hadrons such as elastic scattering, string and resonance excitation and strangeness exchange reactions are implemented \cite{Steinheimer:2018fja}. Many final observables are very sensitive to the hadronic scattering \cite{Stefaniak:2018wwh}. 

\section{The two-particle source}
\label{s:source}

In experimental heavy-ion physics, investigation of quantum-statistical (femtoscopic) correlations is one of the most important methods to gain information about the space-time geometry of the particle emitting source~\cite{Lisa:2005dd,Csorgo:1999sj}. The method of accessing the source function experimentally is to first assume a shape for the source distribution, then calculate the corresponding shape for the two-particle momentum correlations, and finally fit the measured correlations and test the initial assumption. If the theoretically calculated function fits the measured correlations in a statistically acceptable way then one can interpret the fitted source parameters. The connection between the $C_2(k,K)$ two-particle momentum correlations and the $D(r,K)$ pair source function can be written as  
\begin{align}
\label{e:corr}
C_2(k,K) = \int d^4r D(r,K)\big|\Psi^{(2)}_{k}(r)\big|^2,
\end{align}
where $k = (p_1 - p_2)/2$ is the relative momentum, $K = (p_1 + p_2)/2$ is the average momentum, $r$ is the pair-separation four vector, and $\Psi^{(2)}_k(r)$ is the symmetrized pair wave function. The pair source is defined as the auto-correlation of the single particle source functions:
\begin{align}
D(r,K) = \int S(\rho + r/2,K)S(\rho-r/2,K)d^4\rho,
\end{align}
where $\rho$ is the pair center of mass four-vector. Recent experimental findings suggest that in heavy-ion physics L\'evy-stable distributions may play the role of the source distribution~\cite{Adare:2017vig,Kincses:2019czd,Porfy:2019scf}. The definition of the symmetric, centered stable distribution in case of spherical symmetry is the following~\cite{Akkelin:1995gh}:
\begin{equation}
\mathcal L(\boldsymbol r; \alpha, R) =\frac{1}{(2\pi)^3} \int d^3\boldsymbol q e^{i\boldsymbol q\boldsymbol r}e^{-\frac{1}{2}|\boldsymbol qR|^\alpha}.
\label{eq:fx}
\end{equation}
The two parameters that describe such distributions are the index of stability $\alpha$ and the scale-parameter $R$. The $\alpha = 2$ case corresponds to the Gaussian distribution, while in case of $\alpha < 2$ the distribution exhibits a power-law behavior. An important property of such distributions is that they retain the same $\alpha$ parameter under convolution of random variables. If one assumes that the single particle source $S(r)$ is a L\'evy-stable distribution, then from the aforementioned property it follows that the pair-source $D(r)$ also has a L\'evy shape with the same index of stability $\alpha$, but different scale parameter:
\begin{equation}
S(r) = \mathcal L(r; \alpha, R) \Rightarrow D(r) = \mathcal L(r; \alpha, 2^{1/\alpha}R).
\end{equation}

Event generators such as the EPOS model provide direct access to the $D(r,K)$ distribution. In the following we present the first results of the shape analysis of the pair source distribution.

\section{Results}
\label{s:results}
\begin{figure}[h!]
\centerline{
\includegraphics[width=0.85\textwidth]{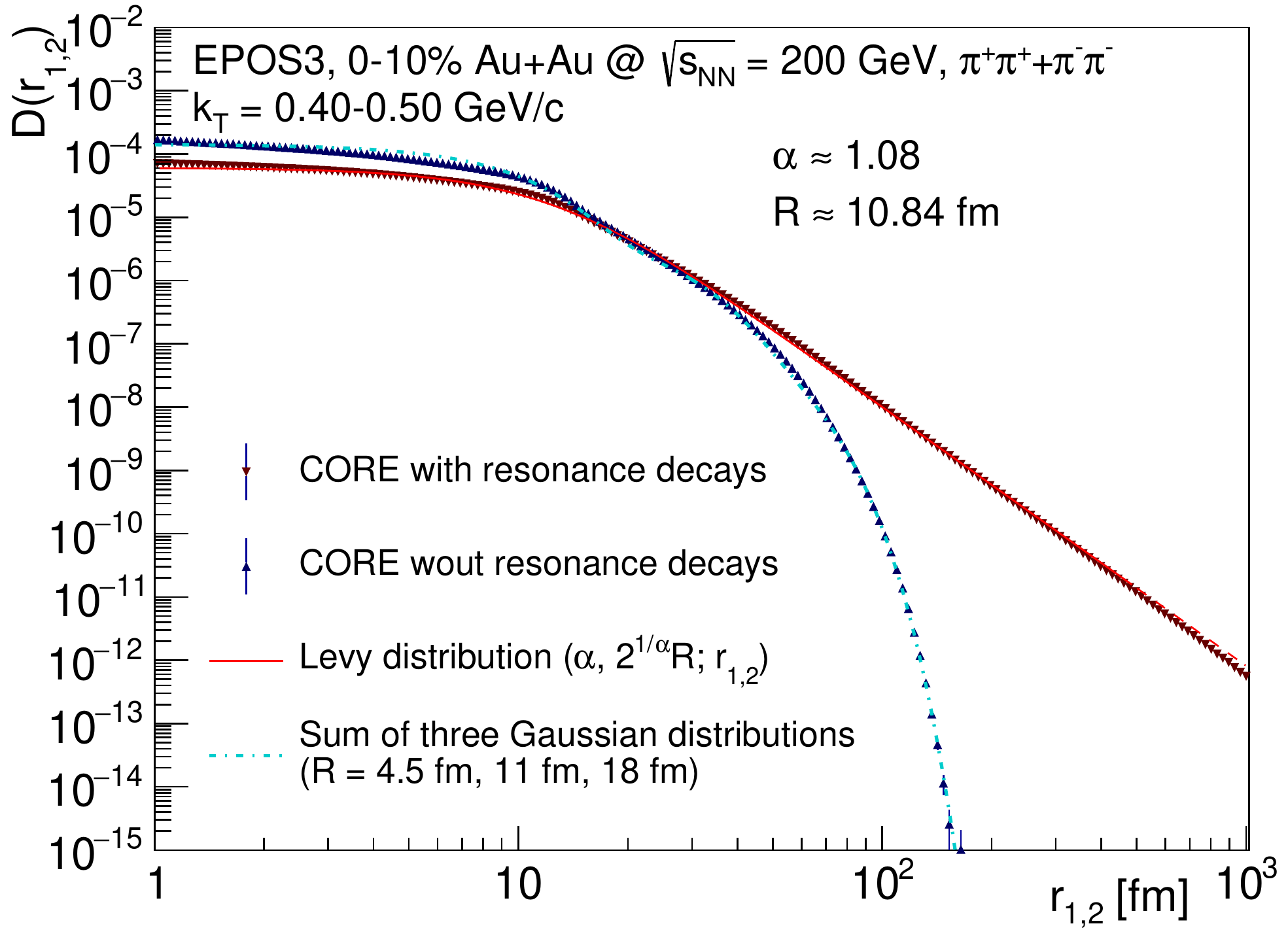}}
\caption{Angle-averaged radial source distribution of like-sign pion pairs generated by EPOS359 for $\sqrt{s_{_{NN}}} = 200$ GeV Au+Au collisions in the centrality range of 0-10\% and average transverse momentum range of 0.4 GeV/$c < k_T <$ 0.5 GeV/$c$. The distribution marked by downward pointing red triangles contains pions created from the hydrodynamically evolving core as well as from the decays of resonances, while the latter is filtered out from the distribution marked by upward pointing blue triangles. The distribution containing resonance decay pions were fitted with a L\'evy distribution in the range of 1 fm - 500 fm as shown with a solid red line. The blue dotted line representing the sum of three Gaussian distributions was not fitted to the corresponding distribution but shown here as an illustration of the approximate trend.\newline}
\label{f:levyfit}
\end{figure}

For the analysis presented below we used $\sqrt{s_{_{NN}}}$ = 200 GeV Au+Au events generated by EPOS359. We kept only particles created from the hydrodynamically expanding core medium, and in our first investigations we omitted the hadronic rescattering. We measured the angle-averaged radial source distribution $D(r_{1,2}) = \int d\Omega dtD(r)$ for like-sign pion pairs in the lab-frame as a function of $r_{1,2} = \sqrt{(x_1-x_2)^2+(y_1-y_2)^2+(z_1-z_2)^2}$ for different ranges of centrality and average transverse momentum $k_T = 0.5\sqrt{k_x^2+k_y^2}$. We applied a single particle transverse momentum cut of 0.15 GeV$/c < p_T < $ 2 GeV$/c$. An example for such measurements can be seen on Fig. \ref{f:levyfit}. 

We investigated two different scenarios of keeping or filtering out resonance decay products from our sample. We observed, that when pions created from decays of resonances are kept in the sample, the shape of the measured distribution closely resembles that of a L\'evy distribution, however, if we remove such pions from our sample the distribution exhibits a Gaussian-like cutoff. While this cutoff is similar to a Gaussian, the whole range in $r_{1,2}$ cannot be described by one such distribution. Without any theoretical assumptions, purely as an illustration of the approximate trend, on Fig. \ref{f:levyfit}. we show that the sum of three different Gaussian distributions can resemble the shape of the measured distribution. In case of keeping the decay products in the sample we fitted the measured distributions with a L\'evy shape in the range of 1 fm to 500 fm. While the confidence level of such fits were not statistically acceptable, the approximate trend can be investigated, and one can extract the centrality and $k_T$ dependence of the two fitted source parameters $\alpha$ and $R$. 

\begin{figure}[h!]
\centerline{\includegraphics[width=0.5\textwidth]{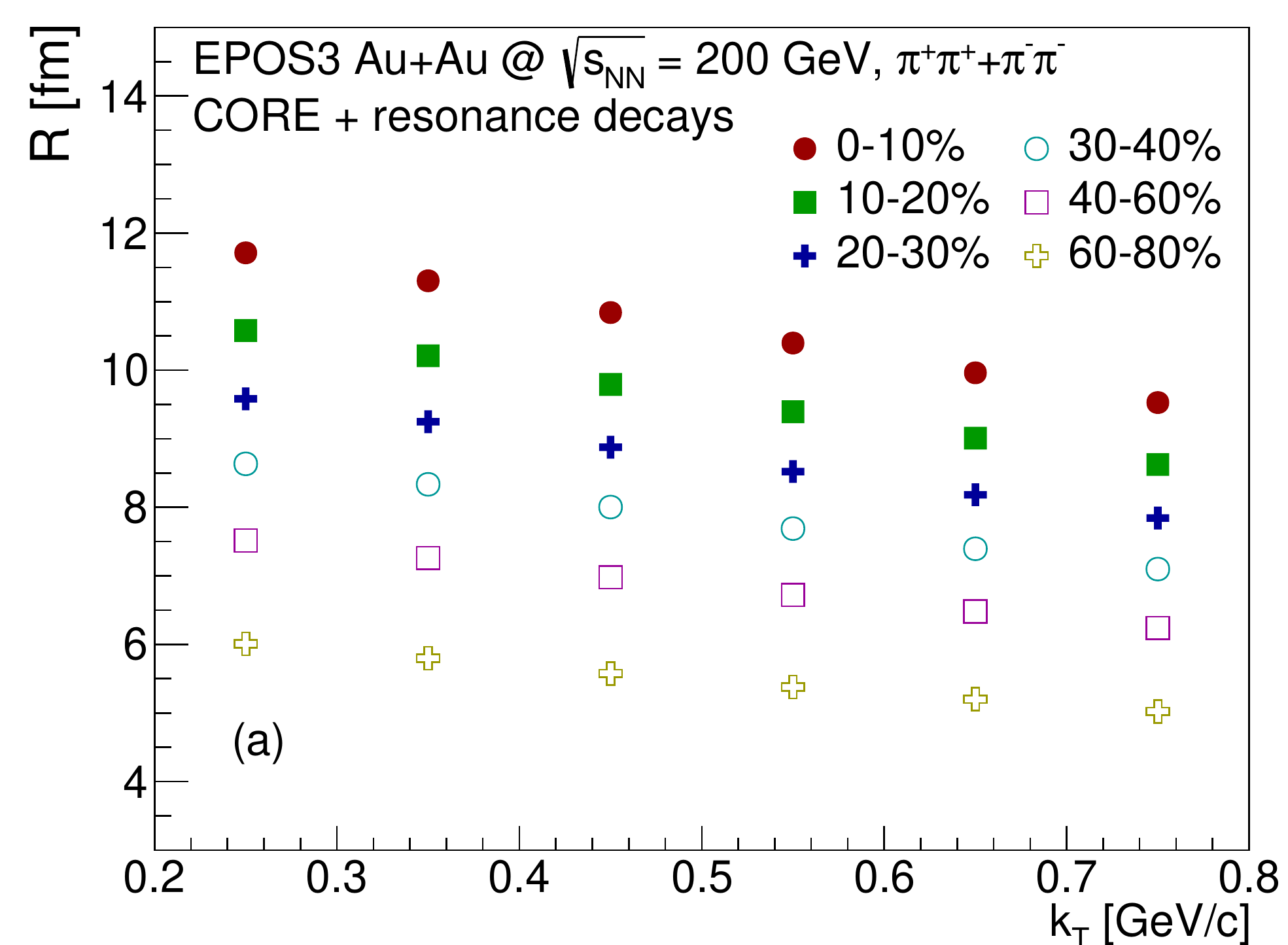}
\includegraphics[width=0.5\textwidth]{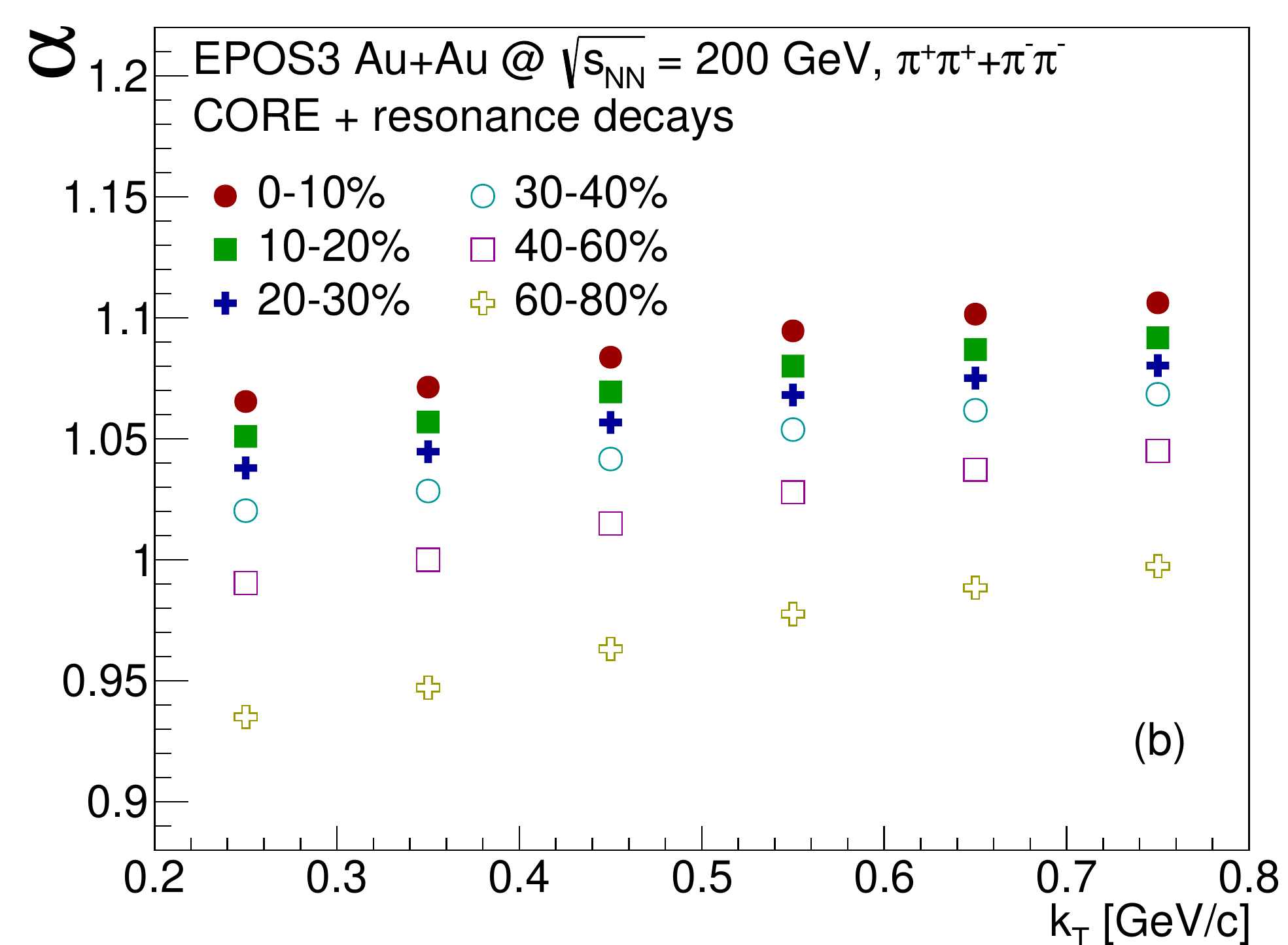}}
\caption{Centrality and average transverse momentum dependence of the extracted L\'evy scale parameter $R$ (a) and the L\'evy exponent $\alpha$ (b) in case of like-sign pion pairs generated by EPOS359 for $\sqrt{s_{_{NN}}} = 200$ GeV Au+Au collisions. }
\label{f:alphaR}
\end{figure}

Fig. \ref{f:alphaR}. (a) shows the centrality and average transverse momentum dependence of the extracted L\'evy scale parameter $R$. The observed trends and magnitudes are quite similar to experimental results~\cite{Adare:2017vig,Lokos:2018dqq}, a geometrical centrality dependence and a decreasing behavior with $k_T$ can be seen. Fig. \ref{f:alphaR}. (b) shows the extracted L\'evy exponent $\alpha$, where an increasing trend can be observed with both $k_T$ and centrality. While the magnitudes are quite similar to experimental results, the trends are different -- in Ref.~\cite{Lokos:2018dqq} it was shown that the $\alpha$ parameter has no strong dependence on $k_T$, and a decreasing trend with centrality was observed. Although in experiment no clear $k_T$ dependence have been observed, other phenomenological studies found similar increasing behavior of the L\'evy exponent~\cite{Tom_ik_2019, Cimerman:2019hva}. The reason behind these apparent contradictions are not clear at this moment. 
\newpage
\section{Summary}
\label{s:sum}

In this paper we presented the first results of an investigation of the pion source function in $\sqrt{s_{_{NN}}} = 200$ GeV Au+Au collisions generated by the EPOS model. We measured the angle averaged radial source distribution of pion pairs as a function of centrality and average transverse momentum. We explored two different scenarios of keeping or removing resonance decay pions from the sample. When removing resonance decays and keeping only particles created from the hydrodynamically evolving core, we observed a Gaussian-like cutoff in the source distribution, however, only one Gaussian fails to describe the whole range in the pair-separation. When keeping resonance decay products in the sample the distribution closely resembles a L\'evy shape. We fitted the measured distributions and extracted the approximate values of the L\'evy source parameters $R$ and $\alpha$. The L\'evy scale $R$ parameter shows the same trend as experimental results, a geometrical centrality dependence, and a decreasing trend with $k_T$ is observed. In case of the L\'evy exponent $\alpha$ parameter we observed similar magnitudes to experimental results, however, the centrality and $k_T$ dependence are in contradiction with experimental findings. Our results show an increasing behavior with both centrality and average transverse momentum, while experimental results indicate no strong dependence on $k_T$ and a decreasing trend with centrality. As of now, the reason behind the opposite behavior is not clear. 

The results discussed in this paper are only preliminary, a more comprehensive analysis is necessary in order to explore the behavior of the two particle source in more details. Our future plans include the investigation of the source distribution in multiple dimensions and different coordinate-frames (e.g. the longitudinal co-moving frame), investigating the effect of different resonance species separately, and also investigating the pair-source of different particles (e.g. kaons, protons). Furthermore, we also plan to conduct a shape analysis on the correlation function reconstructed from the measured source distribution according to Equation \ref{e:corr}.

\section{Acknowledgments}

The authors would like to thank M. Csan\'ad, H. Zbroszczyk and K. Werner for useful discussions. This work was supported by the Grant of National Science Centre, Poland, No: 2017/27/B/ST2/01947. Studies were funded by IDUB-POB-FWEiTE-1 project granted by Warsaw University of Technology under the program Excellence Initiative: Research University (ID-UB). This work was supported by the Polish National Science Center (NCN) under Contracts Nos. UMO-2016/23/B/ST2/01789 and UMO-2017/26/E/ST3/00428. Author D. K. was supported by the Hungarian NKIFH grants No. FK-123842 and FK-123959, as well as the \'UNKP-19-3 New National Excellence Program of the Hungarian Ministry for Innovation and Technology. This work was supported by the COST Action CA15213 THOR.



\bibliographystyle{spiebib} 

\end{document}